\documentclass[aps, prl, superscriptaddress, noeprint, twocolumn, letterpaper,10pt,amsfonts, amsmath, amssymb]{revtex4-1}
\usepackage[utf8]{inputenc}
\usepackage[normalem]{ulem}
\usepackage{bm,physics,graphicx}
\usepackage{hyperref} 
\usepackage{xcolor}
\hypersetup{
    colorlinks,
    linkcolor={blue!80!black},
    citecolor={blue!80!black},
    urlcolor={blue!80!black}
}

\newcommand{\spot}{SP^{\ast}OT^{\ast} = -1}
\newcommand{\req}[1]{Eq.~(\ref{#1})}

\newcommand{\fermif}[1]{n_{\mathrm{F}}\qty(#1)}
\newcommand{\convfac}{e^{i \, \omega_{m}\delta}}
\newcommand{\matsone}{i \, \omega_{m}}
\newcommand{\matstwo}{i \, \omega_{m} + i \, \nu_{n}}
\begin{document}
\title{ Signatures of odd-frequency pairing in the Josephson junction  current noise}
\author{Rub\'{e}n Seoane Souto}
\email{email:ruben.seoane\_souto@ftf.lth.se}
\affiliation{Division of Solid State Physics and NanoLund, Lund University, S-22100 Lund, Sweden}
\affiliation{Center for Quantum Devices and Station Q Copenhagen, Niels Bohr Institute, University of Copenhagen, DK-2100 Copenhagen, Denmark}
\author{Dushko Kuzmanovski}
\affiliation{Nordita KTH Royal Institute of Technology and Stockholm University SE-106 91 Stockholm, Sweden}
\author{Alexander V. Balatsky}
\affiliation{Nordita KTH Royal Institute of Technology and Stockholm University SE-106 91 Stockholm, Sweden}
\affiliation{Department of Physics, University of Connecticut, Storrs, Connecticut 06269, USA}

\date{\today}

\begin{abstract}
Odd frequency (odd-$\omega$) electron pair correlations  naturally appear at the interface between BCS superconductors and other materials. The detection of odd-$\omega$  pairs, which are necessarily non-local in time, is still an open problem. The main reason is that they do not contribute to static measurements described by time-local correlation functions. Therefore, dynamical measurements, which depend on non-local time correlations, are suitable for detecting these pairs. In this work, we study the signatures of odd-$\omega$ pairs in the  supercurrent noise through a weak link between two superconductors at different superconducting phases. We show that the finite frequency current noise can be decomposed into three different contributions coming from even frequency (even-$\omega$), odd-$\omega$ pair amplitudes, and electron-hole correlation functions. Odd-$\omega$ pairing, which is inter-lead (between electrons at different sides of the junction), provides a positive contribution to the noise, becoming maximal at a superconducting phase difference of $\pi$. In contrast, intra-lead even-$\omega$ pair amplitude tends to reduce the noise, except for a region close to $\pi$, controlled by the transmission of the junction.

\end{abstract}
\maketitle
\section{Introduction}

Fermionic wavefunctions  change sign under the exchange of two particles or, equivalently, under the exchange of two electron labels. This leads to the $\spot$ rule~\cite{Balatsky_PRB92,Linder_RMP2019}, which denotes a sign change after the overall exchange of spin ($S$), position ($P^{\ast}$), orbital ($O$) and time ($T^{\ast}$) indexes. In conventional single band ($O = +1$) BCS superconductors, Cooper pairs are spin-singlet ($S=-1$), and even in time and spatial coordinates. In contrast, $p$-wave pairs are spin-triplet pairs ($S = +1$), anti-symmetric under the exchange of spatial coordinates ($P^{\ast} =-1$). In addition to these two possibilities, the $\spot$ rule predicts the existence of electron pairs, which are odd under the exchange of time arguments, i.e., odd-$\omega$ pairs.

Spin-triplet odd-$\omega$ superconductivity was firstly proposed by Berezinskii in the context of $^3$He~\cite{Berezinskii_74Rus, *Berezinskii_74}. Although the existence of these pairs has been ruled out in $^3$He, the idea of generating odd-$\omega$ pairs has driven an intense research activity in the condensed matter community (for recent reviews see Refs.~\cite{Tanaka_JPSJ12,Linder_RMP2019}). It has been demonstrated that electron-electron interactions mediated by phonons cannot stabilize an odd-$\omega$ gap by itself~\cite{Abrahams_PRB93}. In contrast, other mechanisms, such as spin, charge, or current  fluctuations, may be favorable~\cite{Shigeta_PRB11, Shigeta_PRB12,Hugdal_arXiv20}. Disordered systems~\cite{Kirkpatrick_PRL91,Belitz_PRB92}, heavy fermion compounds~\cite{Coleman_PRL93, Coleman_PRB94, Coleman_PRL95, Coleman_JPCM97, Cox_AdvPhis_98}, topological materials \cite{Huang_PRB15,Sukhachov_PRB19,Dushko_PRB20}, and multiband superconductors~\cite{Black_PRB13,Tamura_PRB20,Kanasugi_arXiv20,Tamura_PRB20,Schmidt_PRB20} have been suggested to host odd-$\omega$ pairs in the equilibrium situation. Alternatively, odd-$\omega$ pairs can also appear in driven conventional superconductors, where time translation symmetry is broken~\cite{ Triola_PRB17}.

Another possibility to stabilize odd-$\omega$ pairs is by proximitizing BCS superconductors to other materials, therefore breaking at least one of spatial translation, spin-rotation, or sub-orbital symmetries. At the interface of the two materials, a change on the sign of $S$, $P^{\ast}$, and $O$ labels is accompanied by a sign change on $T^{\ast}$. The conversion between $S$ and $T^{\ast}$ parity may be generated by breaking spin-rotation symmetry either through a Zeeman field in ferromagnet-superconductor interfaces~\cite{Bergeret_PRL01, Bergeret_RMP05}, ferromagnetic insulators~\cite{Linder_ScRep2015}, or systems with spin-orbit coupling~\cite{Black_PRB13_2,Ebisu_PTEP16,Cayao_PRB17,Cayao_PRB18,Dutta_PRB19,Tamura_PRB19_02,Krieger_arXiv2020}. A conversion between the parities of $O$ and $T^{\ast}$ is found at the interface of multiband superconductors,~\cite{Black_PRB13_2,Ebisu_PTEP16, Dushko_PRB17,Dutta_PRB20}. Similarly, a conversion between the parities of $P^{\ast}$ and $T^{\ast}$ is realized in systems where only translation symmetry is broken~\cite{Tanaka_PRB07, Tanaka_PRL07, Tsintzis_PRB19}. 

More recently, the presence of odd-$\omega$ pairs has been predicted to appear at the vicinity of non-magnetic impurities \cite{Kashuba_PRB17,Triola_PRB19}, magnetic impurities \cite{Dushko_PRB2020,Perrin_arXiv19,Santos_arXiv20} and Josephson junctions \cite{Balatsky_arXiv18}. In this work, we revisit a system of this kind, namely two weakly coupled superconductors. In this case, there occurs a conversion between the parity of the lead index ($P
^{\ast}$) and $T^{\ast}$.

The detection of odd-$\omega$ pairs is still an open problem. The difficulty arises from the fact that odd-$\omega$ pairs have vanishing pair amplitude at  equal time. Therefore, static measurements would have to disentangle the dynamic contribution from odd-$\omega$ and conventional contribution to detect the presence of Berezinskii pairs. For this reason, odd-$\omega$ pairing is sometimes also referred as hidden order~\cite{Aeppli_NatRev20}. Instead, dynamic measurements, depending on non-local correlation functions in time, represent  natural probes  to detect the presence of these pairs. In the past, the effects of odd-$\omega$ pairs have been analyzed in the density of states~\cite{Cayao_PRB17,Tsintzis_PRB19,Dushko_PRB2020,Perrin_arXiv19,Santos_arXiv20,Dushko_PRB20} and spectral density~\cite{Komendova_PRB15, Dushko_PRB17, Sukhachov_PRB19, Kanasugi_arXiv20}, Josephson current~\cite{Tanaka_PRL07_2,Balatsky_arXiv18,Dutta_PRB19,Dutta_PRB20}, Meissner effect~\cite{DiBernardo_PRX15,Schmidt_PRB20,Krieger_arXiv2020}, optical conductivity~\cite{Sukhachov_PRB19_02}, spin and charge susceptibilities  ~\cite{Shigeta_PRB12,Higashitani_PRB14,Balatsky_arXiv18} and the current noise through normal leads~\cite{Weiss_PRB17}. Also, spin measurements have been used as an indirect evidence of the presence of odd-$\omega$ pairs~\cite{Bergeret_PRB04,Xia_PRL09}.
Very recently, the Majorana bound states have been proposed as a sensitive probe to the presence of these pairs in conventional superconductors~\cite{Kashuba_PRB17}.

In this work we focus on the supercurrent noise as a possible probe of inter-lead odd-$\omega$ pair correlations.
Current noise was evaluated within the tunnel Hamiltonian approach~\cite{Josephson_RMP64, Ambegaokar_PRL63, *Ambegaokar_PRL63Err, DeGennes_PL63}, as early as the Josephson effect was first proposed~\cite{Josephson_PL62}. However, leading order calculations in the tunneling amplitude only account for intra-lead contributions to the noise. The pair correlations that we are interested in here occur at  all order in the tunneling amplitude, thus, becoming more pronounced at higher transparencies. For high transparent junctions, the multiple Andreev reflections at the interface lead to the onset of the Andreev bound states (ABSs)~\cite{Golubov_RMP04,Martin_Adv2011}. At finite phase difference, they appear inside the superconducting gap, dominating the equilibrium transport properties through the junction. In the past, the supercurrent fluctuations have been extensively analyzed  in Josephson junctions~\cite{Rodero_PRB96,Cuevas_PRL99,Naveh_PRL99,Belzig_PRL01,Cron_PRL01,Cuevas_PRL03,Yip_PRB03,Cuevas_PRB04,Pilgram_PRL05,Galaktionov_PRB10,Ronen_PNAS15,Souto_PRL2016,Souto_PRB2017,Hata_PRL18,Jacquet_EPJ19}.

In this work, we show that the  supercurrent noise in a weak link between two superconductors can be decomposed into three different contributions in the equilibrium situation: a normal contribution, coming from the electron-hole correlation function, an intra-lead even-$\omega$ contribution, which tends to reduce the noise, and an inter-lead odd-$\omega$ contribution, which dominates close to the superconducting phase difference $\pi$. 

The rest of the paper is organized as follows: In Sec.~\ref{sec:Model}, we specify the model and the Green functions formalism for evaluating the expectation values, with the details of the calculation given in the Appendix~\ref{Appendix_NEGF}. In Sec.~\ref{sec:Res}, we demonstrate the results with regards to the pair correlations, and the noise at finite frequencies. The conclusions of our work are summarized in Sec.~\ref{sec:Conc}. An alternative derivation of the current noise from the imaginary-time formalism and using fluctuation-dissipation theorem is presented in Appendix~\ref{appendImagTime}.

\section{\label{sec:Model}Model and formalism}
We consider two bulk superconductors coupled through a weak link, described by the Hamiltonian
\begin{equation}
	H=H_{\mathrm{leads}}+H_{\mathrm{T}}\,,
	\label{Eq:Ham}
\end{equation}
where
\begin{eqnarray}
    & H_{\mathrm{leads}} = \sum_{\nu,k,\sigma} \xi_{\nu}\qty(k) c_{\nu,k\sigma}^\dagger c_{\nu,k\sigma} \nonumber \\
    & + \Delta \, \sum_{\nu,k}(c_{\nu,k\uparrow}^\dagger c_{\nu,-k\downarrow}^\dagger+\mbox{H.c.}) \label{eq:Hamleads}
\end{eqnarray}
with $k$ being the electron momentum, $\sigma$ the spin and $\nu=L,R$. The superconducting gap, $\Delta$, is taken to be the same for both leads for simplicity, and considered real. The tunneling between the two superconductors is described by
\begin{equation}
    H_{\mathrm{T}} =\sum_{k,k'\sigma}\left[V_{k, k'} \, c_{L,k\sigma}^\dagger c_{R,k'\sigma}+\mathrm{H.c.}\right], \label{eq:HamT}
\end{equation}
with $V_{k,k'}$ being the tunneling amplitude. We take the wideband limit, which considers the tunneling amplitude to be energy-independent and described by $V_{k, k'}=V\,e^{i\phi/2}$ with $\phi$ being the superconducting phase difference between the two electrodes. Hereafter we use $\hbar = k_B = e = 1$.

We treat this problem using the non-equilibrium Green functions formalism. In the Keldysh contour, the Green functions are defined as
\begin{equation}
\hat{G}^{\alpha\beta}_{\mu\nu}(t,t')=-i\left\langle \mathcal{T}_c \hat{\Psi}_{\mu,k}(t)\hat{\Psi}_{\nu,k'}^\dagger(t')\right\rangle\,,
\label{Green_Func_def}
\end{equation}
where $\mathcal{T}_c$ is the time order operator in the contour and $\alpha,\beta=\pm$ denote the Keldysh branch for $t$ and $t'$, respectively. We define the Nambu spinor as $\hat{\Psi}_{\nu,k}^\dagger=(c_{\nu,k\uparrow}^\dagger, c_{\nu,-k\downarrow})$ with $\mu,\nu=L,R$ and the hat is used to denote the implicit Nambu structure. The problem can be solved using the retarded and advanced components of the Green functions, which have simple relations to the Keldysh ones. For instance, the time-ordered Green function can be computed as
\begin{equation}
\hat{G}^{++}_{\mu\nu}(\omega)=n_{\mathrm{F}}(\omega) \, \hat{G}^{a}_{\mu\nu}(\omega)+\left[n_{\mathrm{F}}(\omega)-1\right] \, \hat{G}^{r}_{\mu\nu}(\omega)\,,
\label{t_order_GF}
\end{equation}
where $a/r$ denote the advanced/retarded component of the Green function and $n_{\mathrm{F}}\qty(\omega)$ is the Fermi distribution function. The full retarded and advanced Green functions can be obtained by solving the Dyson equation (analytic expressions are given in Appendix \ref{Appendix_NEGF}). These functions have two poles located at $\omega=\pm \epsilon_A$, with $\epsilon_A=\Delta\sqrt{1-\tau\sin^2(\phi/2)}$ being the energy of the ABSs formed at the interface between the two superconductors. Here, $\tau=4V^2/W^2/(1+V^2/W^2)
^2$ is the normal transmission coefficient and $W=1/\pi\rho_F$, where $\rho_F$ is the density of states at the Fermi level.

The current operator can be determined as the time derivative of the electron number operator at one side of the constriction. It is given by
\begin{eqnarray}
	I(t)=i\,V\sum_{k,k'}\mbox{Tr}_\mathrm{N}\left\{\hat{\sigma}_z\left[e^{i\phi/2}\hat{\Psi}_{R,k'}(t) \, \hat{\Psi}_{L,k}^\dagger(t) \right.\right.\nonumber\\
	\left.\left.-e^{-i\phi/2}\hat{\Psi}_{L,k'}(t) \, \hat{\Psi}_{R,k}^\dagger(t)\right]\right\},
	\label{I_def}
\end{eqnarray}
where $\mbox{Tr}_\mathrm{N}$ and $\hat{\sigma}_z$ denote the trace and the Pauli matrix in the Nambu space. In this system, the supercurrent is carried by the ABSs. The mean current has a very compact expression, given by
\begin{equation}
	\left\langle I\right\rangle=\Delta^2\tau\frac{\sin\phi}{2\epsilon_A}\tanh\left(\frac{\epsilon_A}{2 T}\right)\,,
	\label{mean_I_anal}
\end{equation}
where $T$ is the temperature. We note that the mean current, corresponding to taking the mean value in Eq. (\ref{I_def}), only depends on the diagonal Nambu components and, therefore, does not contain information about pairing mechanism in the system. In contrast, the higher order cumulants of the current distribution have terms that depend on the off-diagonal Nambu components of the Green functions evaluated at two different times.  This is essential to have a finite contribution of the odd-$\omega$ pair amplitude.

The current noise spectrum is calculated as
\begin{equation}
    S(\Omega)=\int_{-\infty}^{\infty} \dd{\omega} \left[C(\omega,\Omega)+C(-\omega,\Omega)\right]
    \label{noise_def}
\end{equation}
where $\Omega$ is the frequency. The excess current-current correlation function is given by the Fourier transform of $C(t,t')=\left\langle \delta I(t) \delta I(t')\right\rangle$ with $\delta I(t)=I(t)-\left\langle I(t)\right\rangle$ and the current operator is given in Eq. (\ref{I_def}). In terms of the non-equilibrium Green functions given in the appendix \ref{Appendix_NEGF}, the excess current-current correlation function is given by

\begin{figure}
 \centering
 \includegraphics[width=1\linewidth]{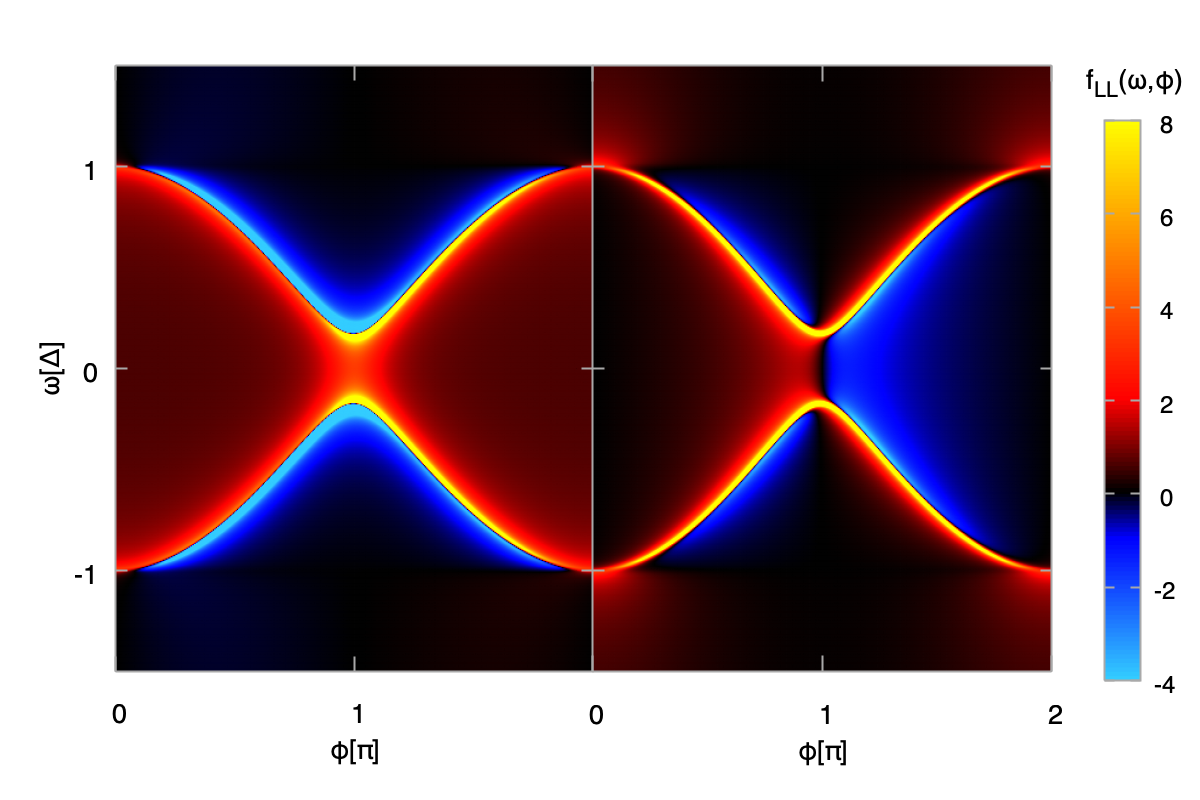}
 \caption{Real (left panel) and imaginary (right panel) part of the intra-lead pair amplitude at the left lead. We take $\tau=0.97$, $T=\Delta/10$ and $\eta=\Delta/100$.}
 \label{fig:Localpairing}
\end{figure}

\begin{eqnarray}
    C(\omega,\Omega)&=&\mbox{Tr}_{\mathrm{N}}\left\{\sum_{\nu=L,R}\hat{\Sigma}_{\nu\bar{\nu}}\left[\hat{G}^{+-}_{\bar{\nu}\nu}(\omega)\hat{\Sigma}_{\nu\bar{\nu}}\hat{G}^{-+}_{\bar{\nu}\nu}(\omega+\Omega)\right.\right.\nonumber\\
    &+&\left.\left.\hat{G}^{+-}_{\bar{\nu}\bar{\nu}}(\omega)\hat{\Sigma}_{\nu\bar{\nu}}\hat{G}^{-+}_{\nu\nu}(\omega+\Omega)\right]\right\}\,,
    \label{I-I_corr}
\end{eqnarray}
where $\bar{\nu}$ denotes the opposite lead to $\nu$. Here $\hat{\Sigma}_{LR}=(\hat{\Sigma}_{RL})^*=V\, \mbox{exp}(i\phi/2 \hat{\sigma}_{z})$. We note that Eq. (\ref{I-I_corr}) contains contributions that depend on the intra-lead and inter-lead pair amplitudes, absent in the current expression (\ref{I_def}), which are analyzed in detail in the next section.

\section{\label{sec:Res}Results}

\subsection{\label{subsec::Pairing}Pair correlation}

\begin{figure}
 \centering
 \includegraphics[width=1\linewidth]{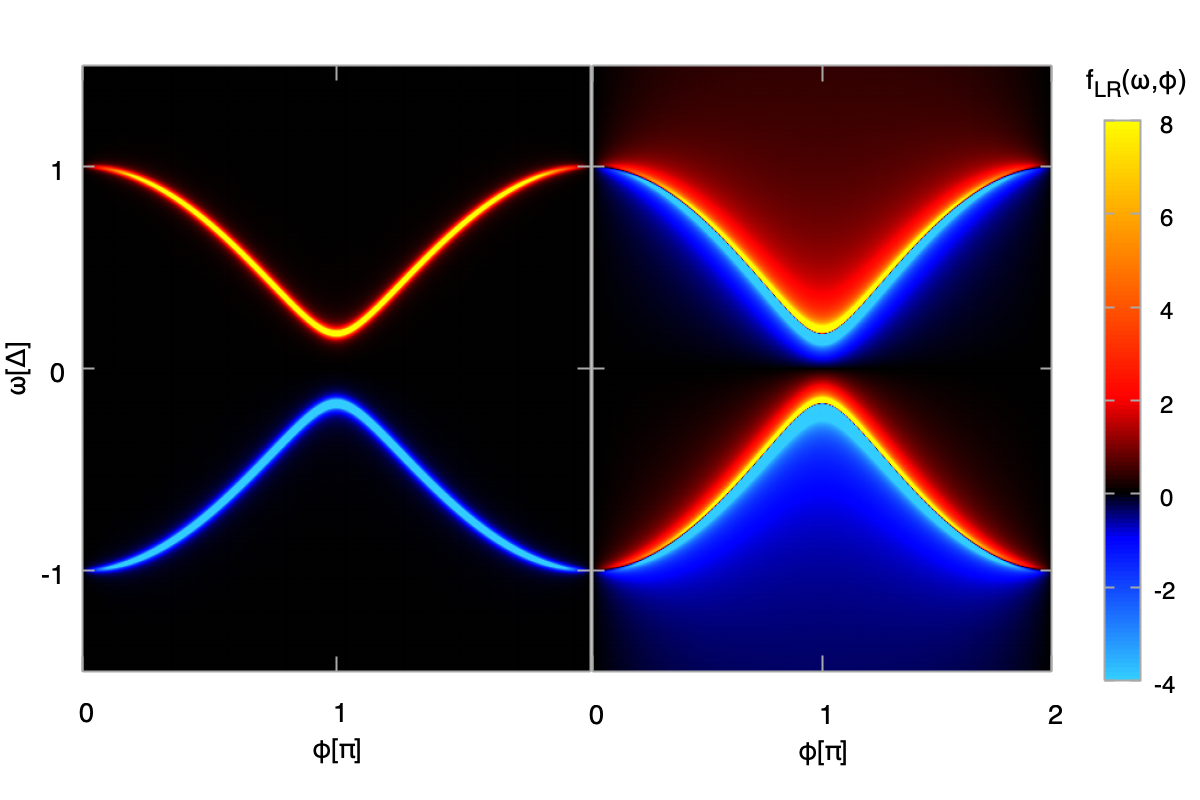}
 \caption{Real (left panel) and imaginary (right panel) part of the inter-lead pair amplitude. The parameters are the same as in Fig. \ref{fig:Localpairing}.}
 \label{fig:non_Localpairing}
\end{figure}

We begin the section by analyzing the pair amplitude between electrons in the junction, defined in the stationary regime as
\begin{eqnarray}
    f_{\mu\nu}(\omega,\phi)=\left[\hat{G}^{++}_{\mu\nu}(\omega)\right]^{12}=\nonumber\\
    -i\sum_{k,k'}\int_{-\infty}^\infty \dd{\omega} e^{i\omega t}\left\langle \mathcal{T} c_{\mu,k\uparrow}(t)c_{\nu,-k'\downarrow}(0)\right\rangle\,,
    \label{pairDef}
\end{eqnarray}
where $\mathcal{T}$ is the causal time order operator, $\eta$ is an infinitesimum and the super-indexes denote the Nambu off-diagonal component of the Green function (\ref{Green_Func_def}). 

We first mention that electron pairs in the junction are spin-singlet ($S=-1$), as there is no mechanism breaking the spin rotation symmetry. The intra-lead pairing at the left electrode is given by
\begin{equation}
    f^{r/a}_{LL}(\omega,\phi)=\frac{\left(1+e^{i\phi}V^2/W^2 \right)\Delta\sqrt{\Delta^2-(\omega\pm i\eta)^2}}{W\left[(\omega\pm i\eta)^2-\epsilon^{2}_A\right](1+V^2/W^2)}\,,
    \label{eq_localPair}
\end{equation}
where the causal component is given by Eq. (\ref{t_order_GF}). The expression in Eq. (\ref{eq_localPair}) is even in frequency, consistent with the $\spot$ rule and $f^{r/a}_{RR}(\omega,\phi)=f^{r/a}_{LL}(\omega,-\phi)$. The pair amplitude is represented in Fig. \ref{fig:Localpairing} for a high transmitting junction. Two sharp features are observed at $\omega=\pm\epsilon_A$ related to the onset of ABSs in the junction. The real part is positive  for $|\omega|<\epsilon_A$, decaying for frequencies bigger than the ABS energy. The imaginary part exhibits a positive peak at $|\omega|=\epsilon_A$ with a contribution at $|\omega|<\epsilon_A$ that changes sign at $\phi=\pi$. Additionally, the imaginary part shows an important contribution for $|\omega|>\Delta$ close to $\phi=2n\pi$, which decays towards $\pi$. In particular, the intra-lead, even-$\omega$ pair amplitude vanishes at $\phi=\pi$ for a perfect transmitting junction,  where the two ABSs merge at zero energy (not shown). This is in agreement with previous calculations showing that even-$\omega$ pair amplitudes are suppressed at the vicinity of a magnetic impurity close to the singlet to doublet ground state transition, where two ABSs meet at zero energy \cite{Dushko_PRB2020}.

The inter-lead pair amplitude is given by 
\begin{equation}
    f^{r/a}_{LR}(\omega,\phi)=\frac{i\tau\Delta\, (\omega\pm i\eta)\sin(\phi/2)}{2V\left[(\omega\pm i\eta)^2-\epsilon^{2}_A\right]}\,,
    \label{odd-w_pair}
\end{equation}
and $f^{r/a}_{RL}(\omega,\phi)=f^{r/a}_{LR}(\omega,-\phi)$. The inter-lead pair amplitude is odd in frequency ($T^\ast=-1$) and anti-symmetric under the exchange of spatial indexes ($P^\ast=-1$), consistent with the $\spot$ rule. Eq. (\ref{odd-w_pair}) is  proportional to the transmission factor, becoming maximal for $\tau=1$. The odd-$\omega$ pair amplitude is represented in Fig. \ref{fig:non_Localpairing}. As shown, it vanishes at $\pi=0$ (mod $2\pi$), where it changes sign, becoming maximal at $\phi=\pi$. We note that the phase of the inter-lead odd-$\omega$ pair amplitude is a $4\pi$ periodic function of $\phi$ (\ref{odd-w_pair}). This in contrast to its modulus which, similarly to the intra-lead pair amplitude (\ref{eq_localPair}), is $2\pi$ periodic. As shown in Fig. \ref{fig:non_Localpairing}, the real part of the odd-$\omega$ contribution has only a finite value at the ABS energy, controlled by the chosen $\eta$ parameter. The imaginary part is also maximum at $|\omega|=\epsilon_A$, showing a finite contribution for $|\omega|>\epsilon_A$. In contrast to $f_{LL}$, the inter-lead pair amplitude does not vanish when the two ABSs merge together for $\tau=1$ and $\phi=\pi$, but it gets enhanced instead.

\subsection{\label{subsec:Noise}Finite frequency noise}

\begin{widetext}

\begin{figure}[htb!]
 \centering
 \includegraphics[width=1\textwidth]{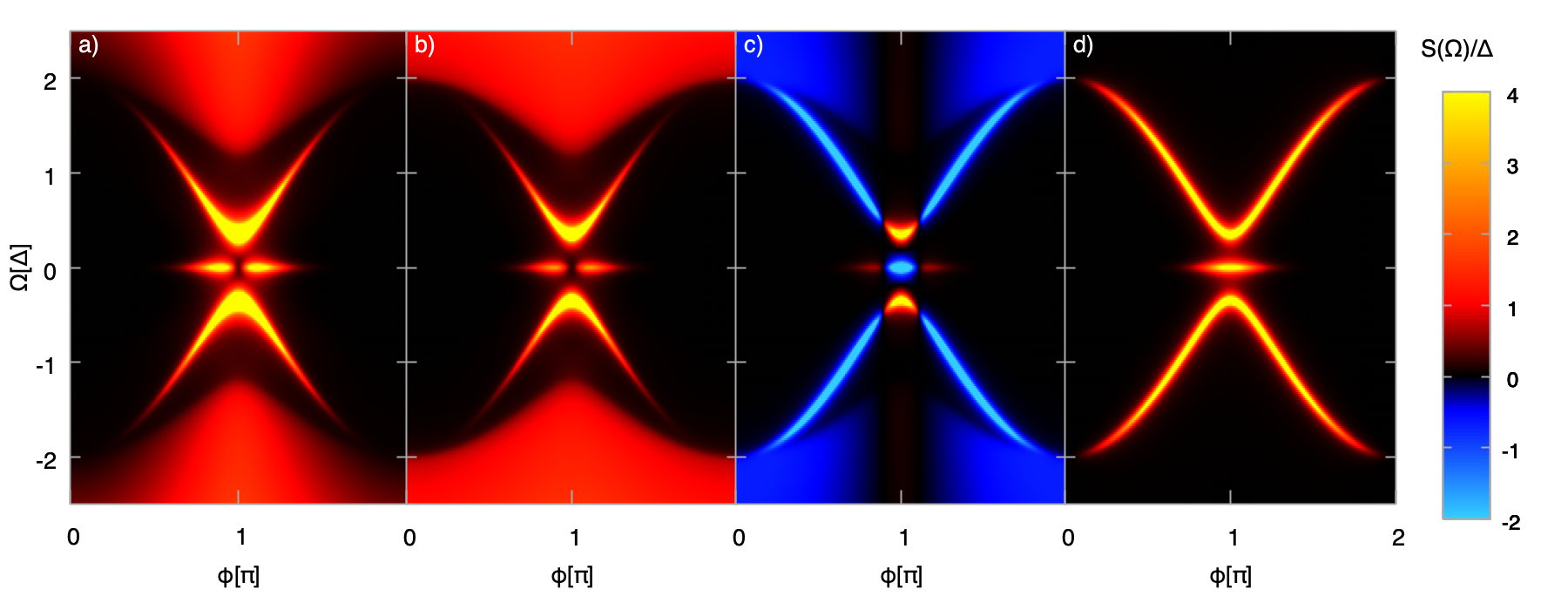}
 \caption{\label{fig:noise-w}Finite frequency noise, where we show: a) Total value, b) electron-hole, c) even-$\omega$ and d) odd-$\omega$ contributions as a function of the phase difference. The parameters are the same as in Fig.~\ref{fig:Localpairing}.}
\end{figure}
\end{widetext}

In this section we analyze the current noise at finite frequency, computed using \req{noise_def}. To analyze the signatures of the different pair amplitudes, we factorize the current noise spectrum into contributions coming from even-$\omega$ and odd-$\omega$ pair amplitudes and  electron-hole correlation functions, which is the only one remaining in the normal state. As commented above, the even-$\omega$  and odd-$\omega$ contributions are given by the Nambu off-diagonal component of the intra- and inter-lead Green function, respectively. Therefore, the excess current-current correlation function in Eq. (\ref{I-I_corr}) can be decomposed as
\begin{widetext}
\begin{subequations}
\begin{eqnarray}
    C_{e}(\omega,\Omega)&=&\sum_{\nu,n}\hat{\Sigma}_{\nu\bar{\nu}}^{nn}\left[\hat{G}^{+-}_{\bar{\nu}\bar{\nu}}(\omega)\right]^{n\bar{n}}\hat{\Sigma}_{\nu\bar{\nu}}^{\bar{n}\bar{n}}\left[\hat{G}^{-+}_{\nu\nu}(\omega+\Omega)\right]^{\bar{n}n}, \\
    C_{o}(\omega,\Omega)&=&\sum_{\nu,n}\hat{\Sigma}_{\nu\bar{\nu}}^{nn}\left[\hat{G}^{+-}_{\bar{\nu}\nu}(\omega)\right]^{n\bar{n}}\hat{\Sigma}_{\nu\bar{\nu}}^{\bar{n}\bar{n}}\left[\hat{G}^{-+}_{\nu\nu}(\omega+\Omega)\right]^{\bar{n}n},
\end{eqnarray}
\end{subequations}
\end{widetext}
corresponding to the even and odd-$\omega$ contributions. Here, the super-index $n=1,2$ runs in the Nambu space and the electron-hole contribution is given by $C_{e-h}(\omega, \Omega)=C(\omega, \Omega)-C_e(\omega, \Omega)-C_o(\omega, \Omega)$.  An alternative representation of each of these components is outlined in Appendix~\ref{appendImagTime} using equilibrium Green functions.

The total current noise is represented in panel a) of Fig. \ref{fig:noise-w} and the three different contributions to the noise are shown in panels b) to d), for a high transmitting junction. We first note that the noise rises up at $|\Omega|>\Delta+\epsilon_A$, corresponding to transitions between the ABSs and the continuum of states. This behavior is dominated by the  electron-hole contribution, which is always positive. In contrast, the even-$\omega$ pair amplitude tends to reduce the current noise for almost any phase difference, except close to $\pi$. It leads to a strong suppression of the noise at $\phi=0$ (mod $2\pi$), where even-$\omega$ pair amplitude becomes maximal for frequencies larger than the superconducting gap, as shown in Fig. \ref{fig:Localpairing}. Close to $\phi=\pi$, the even-$\omega$ contribution becomes smaller exhibiting a sign change, which is better seen in the subgap features. Finally, the contribution from the odd-$\omega$ pair amplitude is negligible for $|\Omega|>\Delta+\epsilon_A$.

\begin{figure}[hbt!]
 \centering
 \includegraphics[width=1\linewidth]{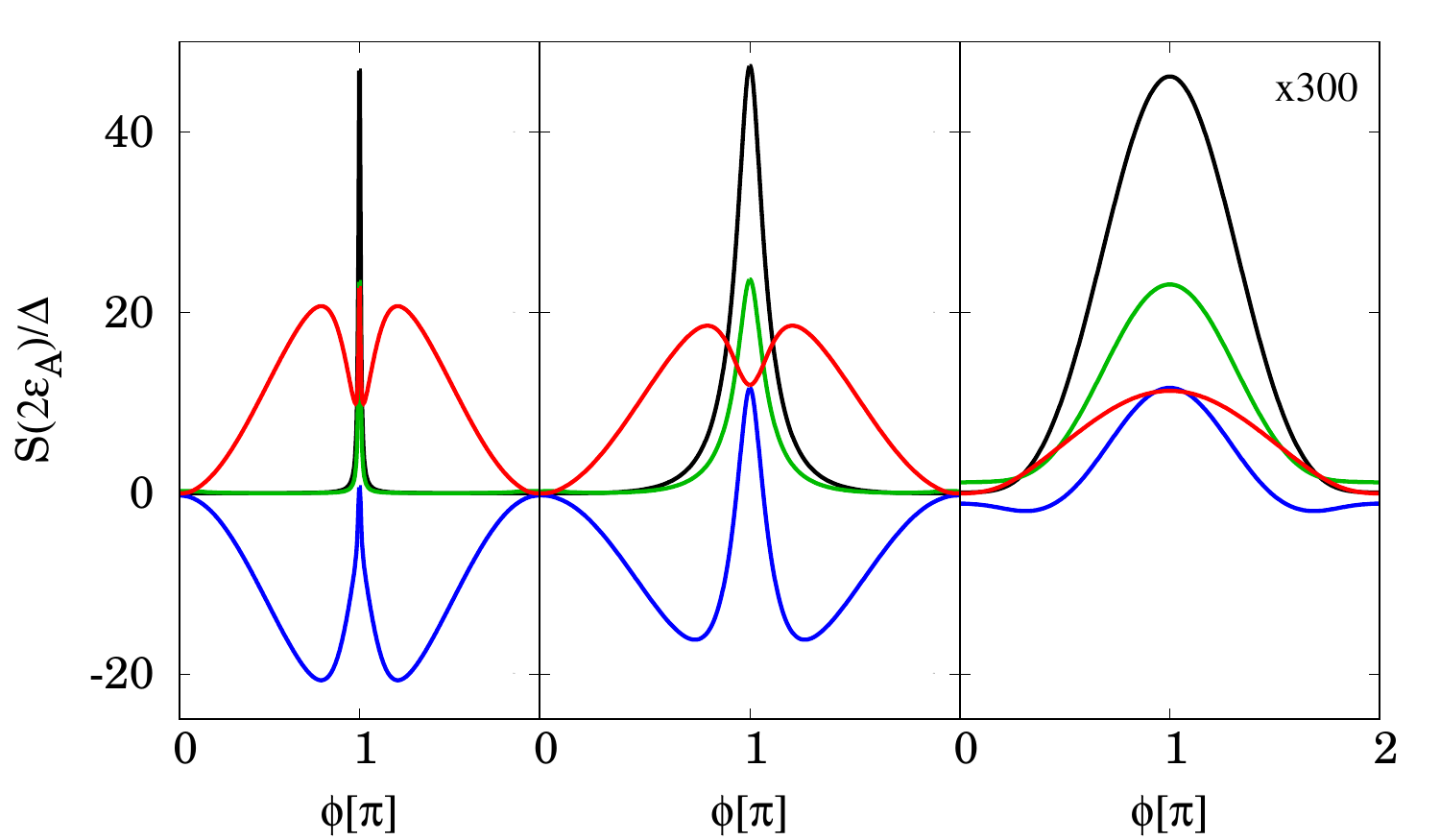}
 \caption{Noise at $\Omega=\pm2\epsilon_A$ (black line), decomposed into electron-hole  (green), even-$\omega$ (blue) and odd-$\omega$ (red). We show cases for $\tau=1$, $0.97$ and $0.1$, from left to right. The lowest transmission case has been scaled up for clarity. The remaining parameters are the same as in Fig. \ref{fig:noise-w}.}
 \label{fig:noise_w2Ea}
\end{figure}

The sharp feature at $\Omega=\pm2\epsilon_A$ due to transitions between the ABSs is analyzed in Fig. \ref{fig:noise_w2Ea} for different transmission factors. For a perfect transmission situation (left panel of Fig. \ref{fig:noise_w2Ea}),  the contributions from  even-$\omega$ and odd-$\omega$ pair amplitudes cancel out, leading to a vanishing noise at almost any phase difference, except close to $\phi=\pi$. At $\phi=\pi$, where the two ABSs merge at zero energy, we observe a peak in the noise which scales up with temperature. At this point, the odd-$\omega$ and the electron-hole contribution are equally important, while the even-$\omega$ one vanishes. 

For a non-perfect, but high transmitting situation (middle panel of Fig. \ref{fig:noise_w2Ea}), we observe a similar cancellation between even and odd-$\omega$ contributions for a wide range of superconducting phases. However, close to $\phi=\pi$ the even-$\omega$ changes sign, enhancing the current noise peak close $\phi=\pi$ for $\tau<1$. The sign change on the even-$\omega$ contribution happens for
\begin{equation}
    \phi_e=\arccos\left(-\tau\right),\quad\mathrm{mod}(2\pi)\,,
    \label{even_crossing}
\end{equation}
providing a positive contribution to the noise for $\phi\in(\phi_e,2\pi-\phi_e)$ (mod $2\pi$).
We note that, for $\Omega=\pm2\epsilon_A$ the sum of even and odd-$\omega$ is equal to the electron-hole contribution for any transmission and phase difference, except close to $\phi=0$, where odd-$\omega$ contribution vanishes and the even-$\omega$ one becomes negative. Therefore, at the points  where the even-$\omega$ vanish given by Eq. (\ref{even_crossing}), the odd-$\omega$ contribution is $1/2$ of the total noise.

\begin{figure}[htb!]
 \centering
 \includegraphics[width=1\linewidth]{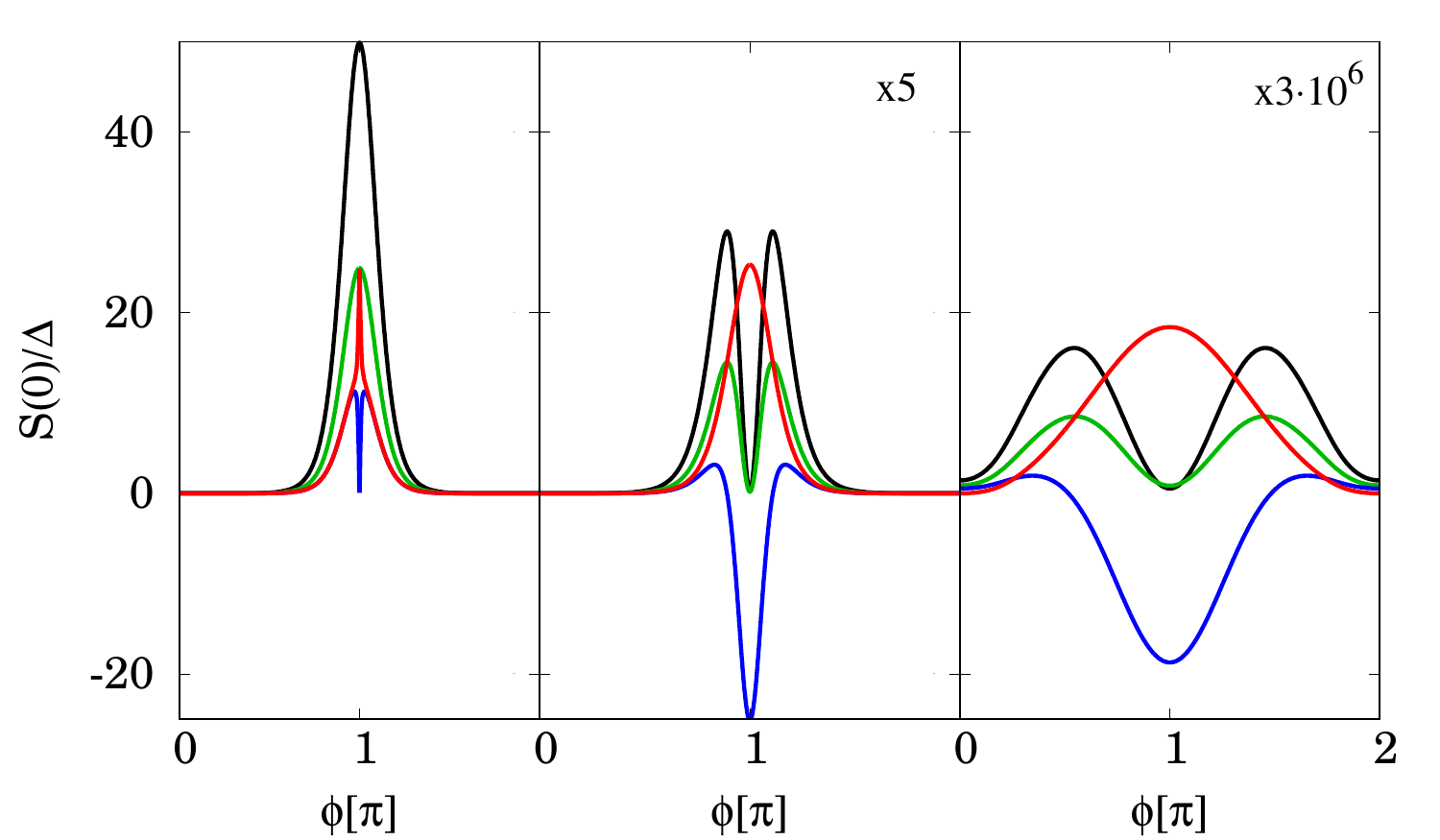}
 \caption{Zero frequency noise (black line), decomposed into electron-hole  (green), even-$\omega$ (blue) and odd-$\omega$ (red). We show cases for $\tau=1$, $0.97$ and $0.1$, from left to right. The remaining parameters are the same as in Fig. \ref{fig:noise-w}.}
 \label{fig:noise_w0}
\end{figure}

For a very low transmission factor (right panel of Fig. \ref{fig:noise_w2Ea}), we observe a strong suppression of the noise, consistent with the result found in Ref. \cite{Rodero_PRB96}. Similarly to the high transmitting situation, the current noise exhibits a peak at $\phi=\pi$ due to the three contributions in the three components. In this limit, the sign change on the even-$\omega$ contribution happens at $\phi=\pi/2$, corresponding to the limit $\tau\to0$ in Eq. (\ref{even_crossing}). 

The current noise at zero frequency is shown in Fig. \ref{fig:noise_w0} for the same parameters as in Fig. \ref{fig:noise_w2Ea}. We note that the zero frequency noise depends on the temperature, vanishing in the $T\to0$ limit. For the perfect transmission situation, the current noise exhibits a peak at $\phi=\pi$. Both electron-hole and odd-$\omega$ contributions are positive for the whole range of superconducting phases, while the even-$\omega$ pair amplitude shows a dip, vanishing at $\phi=\pi$. We note that the sign of the even-$\omega$ component at zero frequency is opposite to the one at $\Omega=\pm2\epsilon_A$ and $|\Omega|>\Delta+\epsilon_A$. For a non-perfect transmitting junction, the current noise exhibits a dip at $\phi=\pi$, due to a cancellation between the even and the odd-$\omega$ contributions. The  maximum of the zero frequency noise is located at the superconducting phase difference given by Eq. (\ref{even_crossing}), where the even-$\omega$ contribution vanishes and the odd-$\omega$ and electron-hole contributions are equally important. In the right panel of Fig. \ref{fig:noise_w0} we show the zero frequency noise for a low transmitting junction, which  exhibits two peaks locate at $\phi=\pi/2+n\pi$, where the  even-$\omega$ contribution vanishes and  the odd-$\omega$ one is $1/2$ of the total noise.

\begin{widetext}

\begin{figure}[htb!]
 \centering
 \includegraphics[width=1\textwidth]{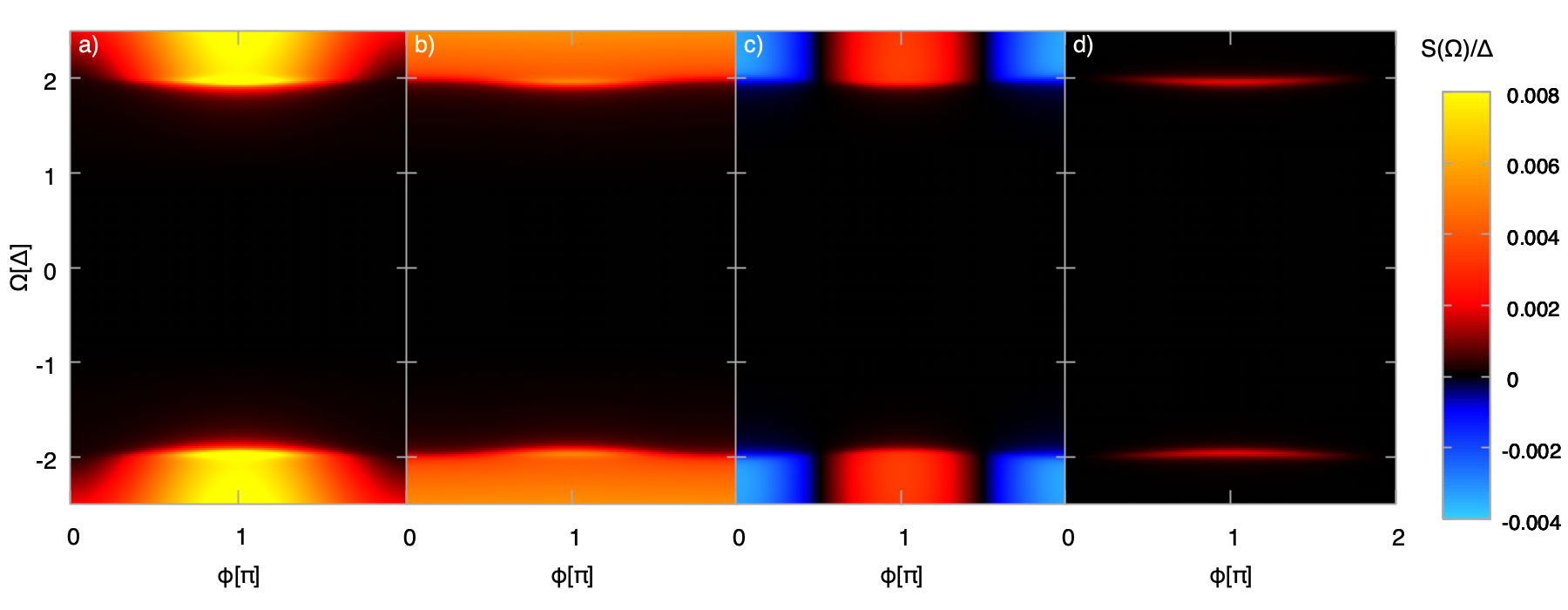}
 \caption{\label{fig:noise_low_tau}Finite frequency noise in the tunnel limit, where we show: a) Total value, b) electron-hole, c) even-$\omega$ and d) odd-$\omega$ contributions as a function of the phase difference. The parameters are the same as in Fig.~\ref{fig:Localpairing}, with $\tau=0.1$.}
\end{figure}
\end{widetext}

Finally, in Fig. \ref{fig:noise_low_tau} we show the current noise in the tunnel limit $\tau\ll1$. In this regime, the ABSs stick at $\epsilon_A=\pm\Delta$, with all the subgap features being suppressed. In this regime, the current noise is dominated by transitions between the continuum states,  $|\Omega|>2\Delta$. The main contribution is provided by the electron-hole correlation functions, represented in panel b), which exhibits a  weak phase dependence. The  supercurrent noise exhibits a maximum at $\phi=\pi$ and a minimum at $\phi=0$. This behavior is controlled by the even-$\omega$ contribution, which changes sign at $\phi=\pi/2+n\pi$, becoming negative for $\phi\in(\pi/2,3\pi/2)$ mod($2\pi$). In the low transmission regime the odd-$\omega$ contribution is suppressed, consistent with the fact that odd-$\omega$ pair amplitude scales linearly with the transmission factor according to Eq. (\ref{odd-w_pair}).

\section{\label{sec:Conc}Conclusions}
In this work, we have shown the onset of odd-$\omega$ electron pairs at the interface between two superconductors at different phases. They are spin-singlet pairs formed by electrons at different superconducting leads. While they do not contribute to the Josephson current, they provide an important contribution to the current noise for high transmitting junctions. In certain regimes, the contribution of the odd-$\omega$ component to the noise is as large as the sum of the other contributions. 
To explore relative strength of different channels we factorized the  equilibrium current noise   into contributions from electron-hole correlation functions and even-$\omega$ and odd-$\omega$ pair amplitudes. 
The odd-$\omega$ contribution is always positive, similarly to the electron-hole, becoming maximal at a superconducting phase difference $\phi=\pi$. In contrast, the even-$\omega$ contribution tends to reduce the current noise for a wide range of superconducting phase differences, except for a range close to $\pi$, controlled by the transmission factor. Our results pave the way to reveal   the odd-$\omega$ Berezinskii pairing through the current noise. One would need to measure phase-biased junctions to probe a strong contribution from the odd-$\omega$  in Josephson junctions.  

This relatively simple picture can be modified by self-consistent effects, which lead to the appearance of several solutions close to a phase difference $\pi$~\cite{Rodero_PRL94,Golubov_RMP04}, also inducing intra-lead odd-$\omega$ pair amplitudes.

\begin{acknowledgments}
The authors wish to thank Pavlo O.~Sukhachov and A. Levy Yeyati for useful discussions. D.K. and A.V.B. acknowledge funding by VILLUM FONDEN via the Centre of Excellence for Dirac Materials (Grant No. 11744),University of Connecticut, and the European Research Council ERC-2018-SyG HERO. R.S.S. acknowledges funding from Nanolund and from QuantERA project 2D hybrid materials as a platform for topological quantum computing.
\end{acknowledgments}

\bibliographystyle{apsrev4-1}

\bibliography{main.bib}

\appendix

\section{\label{Appendix_NEGF}Green functions formalism}
The equilibrium Green functions of the bare electrodes are described by \cite{Martin_Adv2011}
\begin{equation}
	\hat{g}^{r/a}_{\mu\nu}(\omega)=\frac{\delta_{\mu\nu}}{W\sqrt{\Delta^2-(\omega\pm i\eta)^2}}\begin{pmatrix}
-\omega\pm i\eta & \Delta\\
\Delta & -\omega\pm i\eta
\end{pmatrix}\,,
\end{equation}
where $\eta$ is an infinitesimal to avoid the divergence of the Green functions at $\omega=\eta$. Here,  $\mu,\nu=L,R$ and the hat denotes the Nambu space. In this space, the tunneling matrix is given by
\begin{equation}
	\hat{V}_{LR}=V\begin{pmatrix}
e^{i\phi/2} & 0\\
 0 & -e^{-i\phi/2}
\end{pmatrix}\,,
\end{equation}
and $\hat{V}_{RL}=\hat{V}_{LR}^*$. In order to obtain the inter-lead Green functions, which determine the current through the system, we solve the Dyson equation, containing information about all-order tunneling processes between the electrodes. In frequency space, it is given by
\begin{equation}
	\hat{G}^{r/a}_{\mu\nu}(\omega)=\hat{g}^{r/a}_{\mu\nu}(\omega)\hat{\delta}_{\mu\nu}+\sum_{\alpha=L,R}\hat{g}^{r/a}_{\mu\alpha}(\omega)\hat{V}_{\alpha\bar{\alpha}}\hat{G}^{r/a}_{\bar{\alpha}\nu}(\omega)\,,
\end{equation}
were the bar over the lead subindex is used to denote the opposite lead. The solution to the equation for the inter-lead components is given by
\begin{equation}
	\hat{G}^{r/a}_{LR}(\omega)=\hat{V}_{RL}^{-1}\tilde{G}^{r/a}_{LR}(\omega)-\hat{V}_{RL}^{-1}\,,
	\label{G_LR}
\end{equation}
where correlation effects are described by the auxiliary function $\tilde{G}^{r/a}_{LR}=(\hat{1}-\hat{V}_{RL}\cdot\hat{g}^{r/a}_{L}\cdot\hat{V}_{LR}\cdot\hat{g}^{r/a}_{R})^{-1}$. Its expression is given by
\begin{widetext}
\begin{equation}
	\tilde{G}^{r/a}_{LR}(\omega)=\frac{1}{\left[(\omega\pm i\eta)^2-\epsilon^{2}_A\right](1+x^2)^2}
	\begin{pmatrix}
		(1+x^2)(\omega\pm i\eta)^2-\Delta^2(1+e^{i\phi}x^2) & x^2\Delta (\omega\pm i\eta) (e^{-i\phi}-1)\\
		x^2\Delta (\omega\pm i\eta) (e^{i\phi}-1)& (1+x^2)(\omega\pm i\eta)^2-\Delta^2(1+e^{-i\phi}x^2)
	\end{pmatrix}
\end{equation}
\end{widetext}
where $x=V/W$ and
\begin{equation}
	\epsilon_A({\phi})=\pm \Delta\sqrt{1-\tau\sin^2(\phi/2)}\,,
	\label{ABSs_pos}
\end{equation}
is the energy of the Andreev bound states (ABSs) with $\tau=4x^2/(1+x^2)^2$. 

In a similar way 
\begin{equation}
	\hat{G}^{r/a}_{RL}(\omega)=\hat{V}_{LR}^{-1}\tilde{G}^{r/a}_{RL}(\omega)-\hat{V}_{LR}^{-1}\,,
	\label{G_RL}
\end{equation}
with $\tilde{G}_{RL}^{r/a}=(\tilde{G}_{LR}^{a/r})^*$. 

The intra-lead Green functions at each side of the junction can be obtained through the Dyson Equation, finding
\begin{eqnarray}
	\hat{G}^{r/a}_{\nu\nu}(\omega)=\tilde{G}^{r/a}_{\mu\nu}(\omega)\cdot\hat{g}^{r/a}_{\nu\nu}(\omega)\,.
	\label{G_LL}
\end{eqnarray}
The Keldysh Green function can be obtained in the equilibrium situation as 
\begin{subequations}
\begin{eqnarray}
	\hat{G}^{+-}_{\mu\nu}(\omega)&=&\fermif{\omega}\left[\hat{G}^{a}_{\mu\nu}(\omega)-\hat{G}^{r}_{\mu\nu}(\omega)\right], \\
	\hat{G}^{-+}_{\mu\nu}(\omega)&=&\left[\fermif{\omega}-1\right]\left[\hat{G}^{a}_{\mu\nu}(\omega)-\hat{G}^{r}_{\mu\nu}(\omega)\right],
\end{eqnarray}
\end{subequations}
where $\fermif{\omega}$  is the Fermi distribution function. 

The current can be computed through the time evolution of the electron number operator in one side
\begin{equation}
    I\qty(t) = i \, \qty[H_{\mathrm{T}}, N_{L}].
\end{equation}
Using \req{eq:HamT} and $N_{L} = \sum_{k,\sigma} c^{\dagger}_{L, k\sigma} c_{L,k\sigma}$, we get the following expression for the current operator
\begin{equation}
    \label{eq:CurrentOp}
    I(t) =V \sum_{k,k',\sigma} \Bqty{ie^{i\phi}  c^{\dagger}_{L,k\sigma}(t) \, c_{R,k', \sigma}(t) + \mathrm{H.c.}}.
\end{equation}
The summation over spin may be traded for a summation over holes, giving rise to the expression \req{I_def}, whose mean value is given in terms of the Keldysh Green functions as
\begin{equation}
	\left\langle I\right\rangle=\mbox{Tr}_{\mathrm{N}}\left\{\hat{\sigma}_z\int \dd{\omega} \left[\hat{V}_{LR}\hat{G}_{RL}^{+-}(\omega)-\hat{V}_{RL}\hat{G}_{LR}^{+-}(\omega)\right]\right\}\,.
	\label{mean_I}
\end{equation}
Integrating the expression, and noticing that the contribution outside the gap vanishes, we obtain the expression in Eq. (\ref{mean_I_anal})

\section{\label{appendImagTime}Expression for current noise in the imaginary-time formalism}
Using \req{eq:CurrentOp}, we can evaluate the current-current susceptibility in the imaginary time formalism
\begin{widetext}
\begin{eqnarray*}
    & \chi\qty(t, t') = -\expval{T_{t} \delta I\qty(t) \, \delta I \qty(t')} \\
    & = \sum_{k, k' p, p' \alpha, \beta} \left\lbrace V_{k, k'} \, V_{p, p'} \expval{T_{t} : \bar{c}_{L, k\alpha}\qty(t + \delta) \, c_{R, k'\alpha}\qty(t)::\bar{c}_{L, p\beta}\qty(t' + \delta) \, c_{R, p'\beta}\qty(t'):} \right. \\
    & -V_{k, k'} \, V^{\ast}_{p, p'} \expval{T_{t} : \bar{c}_{L, k\alpha}\qty(t + \delta) \, c_{R, k'\alpha}\qty(t)::\bar{c}_{R, p\beta}\qty(t' + \delta) \, c_{L, p'\beta}\qty(t'):} \\
    & -V^{\ast}_{k, k'} \, V_{p, p'} \expval{T_{t} : \bar{c}_{R, k\alpha}\qty(t + \delta) \, c_{L, k'\alpha}\qty(t)::\bar{c}_{L, p\beta}\qty(t' + \delta) \, c_{R, p'\beta}\qty(t'):} \\
    & \left. +V^{\ast}_{k, k'} \, V^{\ast}_{p, p'} \expval{T_{t} : \bar{c}_{R, k\alpha}\qty(t + \delta) \, c_{L, k'\alpha}\qty(t)::\bar{c}_{R, p\beta}\qty(t' + \delta) \, c_{L, p'\beta}\qty(t'):}\right\rbrace,
\end{eqnarray*}
where $\delta$ is a small time displacement and the semicolons indicate that the pairing terms should be discarded in evaluating $\delta I\qty(t)$. Applying Wick's Theorem to the $4$-point imaginary-time-ordered correlation functions, and using the fact that for spin-singlet pairing and time-independent potentials the $2$-point correlators are given by
\begin{subequations}
\label{eq:TwoPointCorr}
\begin{eqnarray}
    & \expval{T_{t} \, c_{\nu, k\alpha}(t) \, \bar{c}_{\nu', k'\beta}\qty(t')} = -\delta_{\alpha\beta} \, G_{\nu\nu'}\qty(k, k'; t - t'), \label{eq:Gee} \\
    & \expval{T_{t} \, c_{\nu, k\alpha}(t) \, c_{\nu', k'\beta}\qty(t')} = -\qty(i \, \sigma_{y})_{\alpha\beta} \, F_{\nu\nu'}\qty(k, k'; t - t'), \label{eq:Geh} \\
    & \expval{T_{t} \, \bar{c}_{\nu, k\alpha}(t) \, \bar{c}_{\nu', k'\beta}\qty(t')} = \qty(i \, \sigma_{y})_{\alpha\beta} \, \bar{F}_{\nu\nu'}\qty(k, k'; t - t'). \label{eq:Ghe}
\end{eqnarray}
\end{subequations}
The current-current susceptibility is given by

\begin{eqnarray}
    & \chi\qty(t - t') = 2 \, \sum_{k, k', p, p'} \left\lbrace V_{k, k'} \, V_{p, p'} \, \left[-G_{RL}\qty(k', p; t - t' - \delta) \, G_{RL}\qty(p', k; t' - t - \delta) \right. \right. \nonumber \\
    & \left. + F_{RR}\qty(k', p'; t - t') \, \bar{F}_{LL}\qty(p, k; t - t')\right] \nonumber \\
    & + V_{k, k'} \, V^{\ast}_{p, p'} \, \left[G_{RR}\qty(k', p; t - t' - \delta) \, G_{LL}\qty(p', k; t' - t - \delta) \right. \nonumber \\
    & \left. - F_{RL}\qty(k', p'; t - t') \, \bar{F}_{RL}\qty(p, k; t - t')\right] \nonumber \\
    & + V^{\ast}_{k, k'} \, V_{p, p'} \, \left[G_{LL}\qty(k', p; t - t' - \delta) \, G_{RR}\qty(p', k; t' - t - \delta) \right. \nonumber \\
    & \left. - F_{LR}\qty(k', p'; t - t') \, \bar{F}_{LR}\qty(p, k; t - t')\right] \nonumber \\
    & + V^{\ast}_{k, k'} \, V^{\ast}_{p, p'} \, \left[-G_{LR}\qty(k', p; t - t' - \delta) \, G_{LR}\qty(p', k; t' - t - \delta) \right. \nonumber \\
    & \left.\left. + F_{LL}\qty(k', p'; t - t') \, \bar{F}_{RR}\qty(p, k; t - t')\right] \right\rbrace. \label{eq:ChiImagTime}
\end{eqnarray}
Performing a Fourier transform $\chi\qty(i \, \nu_{n}) = \int_{0}^{1/T} \dd{t} e^{i \, \nu_{n} t} \, \chi\qty(t)$, where $\nu_{n} = 2 n \pi \, T$ and expressing the Green's functions in terms of fermionic Matsubara frequencies, \req{eq:ChiImagTime} may be rewritten as
\begin{eqnarray}
    & \chi\qty(i \, \nu_{n}) = 2 T \, \sum_{\omega_{m}, k, k', p, p'} \left\lbrace V_{k, k'} \, V_{p, p'} \, \left[-\convfac \, G_{RL}\qty(k', p; \matstwo) \, G_{RL}\qty(p', k; \matsone) \right. \right. \nonumber \\
    & \left. + F_{RR}\qty(k', p'; \matstwo) \, \bar{F}_{LL}\qty(p, k; \matsone)\right] \nonumber \\
    & + V_{k, k'} \, V^{\ast}_{p, p'} \, \left[\convfac \, G_{RR}\qty(k', p; \matstwo) \, G_{LL}\qty(p', k; \matsone) \right. \nonumber \\
    & \left. - F_{RL}\qty(k', p'; \matstwo) \, \bar{F}_{RL}\qty(p, k; \matsone)\right] \nonumber \\
    & + V^{\ast}_{k, k'} \, V_{p, p'} \, \left[\convfac \, G_{LL}\qty(k', p; \matstwo) \, G_{RR}\qty(p', k; \matsone) \right. \nonumber \\
    & \left. - F_{LR}\qty(k', p'; \matstwo) \, \bar{F}_{LR}\qty(p, k; \matsone)\right] \nonumber \\
    & + V^{\ast}_{k, k'} \, V^{\ast}_{p, p'} \, \left[-\convfac \, G_{LR}\qty(k', p; \matstwo) \, G_{LR}\qty(p', k; \matsone) \right. \nonumber \\
    & \left.\left. + F_{LL}\qty(k', p'; \matstwo) \, \bar{F}_{RR}\qty(p, k; \matsone)\right] \right\rbrace. \label{eq:ChiFourier}
\end{eqnarray}

Next, we use the spectral representation of the Green's functions
\begin{subequations}
\label{eq:SpecRep}
\begin{eqnarray}
    & G_{\nu \nu'}(k, k'; z) = \int_{-\infty}^{\infty} \dd{\omega'} \frac{A_{\nu \nu'}\qty(k, k'; \omega')}{z - \omega'}, \label{eq:SpecGee} \\
    & F_{\nu \nu'}(k, k'; z) = \int_{-\infty}^{\infty} \dd{\omega'} \frac{B_{\nu \nu'}\qty(k, k'; \omega')}{z - \omega'}, \label{eq:SpecGeh} \\
    & \bar{F}_{\nu \nu'}(k, k'; z) = \int_{-\infty}^{\infty} \dd{\omega'} \frac{\bar{B}_{\nu \nu'}\qty(k, k'; \omega')}{z - \omega'}, \label{eq:SpecGhe}
\end{eqnarray}
\end{subequations}
and the sum over Matsubara frequencies
\begin{equation}
    \label{eq:MatsSumOne}
    T \sum_{\omega_{m}} \frac{1}{\qty(\matstwo - \omega_{2}) \, \qty(\matsone - \omega_{1})} = \frac{\fermif{\omega_{1}} - \fermif{\omega_{2}}}{i \, \nu_{n} - \omega_{2} + \omega_{1}},
\end{equation}
to re-express \req{eq:ChiFourier} as a double integral over real frequencies with a common denominator.  The only factor that contains the imaginary Matsubara frequency, $i \, \nu_{n}$, is described by
\begin{eqnarray}
    & \chi(i \, \nu_{n}) = \sum_{k, k', p, p'} \int_{-\infty}^{\infty} \int_{-\infty}^{\infty} \dd{\omega_{2}} \dd{\omega_{1}} \frac{\fermif{\omega_{1}} - \fermif{\omega_{2}}}{i \, \nu_{n} - \omega_{2} + \omega_{1}} \nonumber \\
    & \times \left\lbrace V_{k, k'} \, V_{p, p'} \, \qty[-A_{RL}\qty(k', p; \omega_{2}) \, A_{RL}\qty(p', k; \omega_{1}) + B_{RR}\qty(k', p'; \omega_{2}) \, \bar{B}_{LL}\qty(p, k; \omega_{1})]\right. \nonumber \\
    & + V_{k, k'} \, V^{\ast}_{p, p'} \, \qty[A_{RR}\qty(k', p; \omega_{2}) \, A_{LL}\qty(p', k; \omega_{1}) - B_{RL}\qty(k', p'; \omega_{2}) \, \bar{B}_{RL}\qty(p, k; \omega_{1})] \nonumber \\
    & + V^{\ast}_{k, k'} \, V_{p, p'} \, \qty[A_{LL}\qty(k', p; \omega_{2}) \, A_{RR}\qty(p', k; \omega_{1}) - B_{LR}\qty(k', p'; \omega_{2}) \, \bar{B}_{LR}\qty(p, k; \omega_{1})] \nonumber \\
    & \left. + V^{\ast}_{k, k'} \, V^{\ast}_{p, p'} \, \qty[-A_{LR}\qty(k', p; \omega_{2}) \, A_{LR}\qty(p', k; \omega_{1}) + B_{LL}\qty(k', p'; \omega_{2}) \, \bar{B}_{RR}\qty(p, k; \omega_{1})]\right\rbrace. \label{eq:ChiFourierSpec}
\end{eqnarray}
Using the spectral representation for the susceptibility as a function of a complex frequency
\begin{equation}
    \chi\qty(z) = \int_{-\infty}^{\infty} \frac{\dd{\Omega}}{\pi} \, \frac{\rho\qty(\Omega)}{z - \Omega}, \label{eq:ChiSpecRep}
\end{equation}
and comparing with the expression \req{eq:ChiFourierSpec}, we can verify that the expression for $\rho\qty(\Omega)$ amounts only to substituting
\[
\frac{\fermif{\omega_{1}} - \fermif{\omega_{2}}}{i \, \nu_{n} - \omega_{2} + \omega_{1}} \rightarrow \pi \, \delta\qty(\Omega - \omega_{2} + \omega_{1}) \, \qty[\fermif{\omega_{1}} - \fermif{\omega_{2}}].
\]
Furthermore, using the Fluctuation-Dissipation Theorem to relate the noise power $S\qty(\Omega)$ with the spectral density $\rho\qty(\Omega)$
\begin{equation}
    \label{eq:FDT}
    S\qty(\Omega) = \coth\left(\frac{\Omega}{2 T}\right) \, \rho\qty(\Omega)
\end{equation}
The expression can be simplified using hyperbolic trigonometric identities, finding
\[
\coth\qty(\frac{\Omega}{2 T}) \, \qty[ \fermif{\omega_{1}} - \fermif{\omega_{2}}] \, \delta(\Omega - \omega_{2} + \omega_{1}) = \frac{1}{2} \, \qty[1 - \tanh\qty(\frac{\omega_{1}}{2 T}) \, \tanh\qty(\frac{\omega_{2}}{2 T})] \, \delta\qty(\Omega - \omega_{2} + \omega_{1})
\]
we get the following expression for the excess current-current correlation function

\begin{eqnarray}
    & C_{e-h}\qty(\omega,\Omega) = \frac{\pi}{2} \, \sum_{k, k', p, p'} \qty[1 - \tanh\qty(\frac{\omega + \Omega}{2 T}) \, \tanh\qty(\frac{\omega}{2 T})] \nonumber \\
    & \times \left[ -V_{k, k'} \, V_{p, p'} \, A_{RL}\qty(k', p; \omega + \Omega) \, A_{RL}\qty(p', k; \omega) \right. 
     + V_{k, k'} \, V^{\ast}_{p, p'} \, A_{RR}\qty(k', p; \omega + \Omega) \, A_{LL}\qty(p', k; \omega)  \nonumber \\
    & + V^{\ast}_{k, k'} \, V_{p, p'} \, A_{LL}\qty(k', p; \omega + \Omega) \, A_{RR}\qty(p', k; \omega) 
     \left. - V^{\ast}_{k, k'} \, V^{\ast}_{p, p'} \, A_{LR}\qty(k', p; \omega + \Omega) \, A_{LR}\qty(p', k; \omega_{1})\right],\label{eq:ChiFourierSpec_ph}\\
     & C_{e}\qty(\omega,\Omega) =\frac{\pi}{2} \, \sum_{k, k', p, p'} \qty[1 - \tanh\qty(\frac{\omega + \Omega}{2 T}) \, \tanh\qty(\frac{\omega}{2 T})] \nonumber \\
     & \left[V_{k, k'} \, V_{p, p'} \, B_{RR}\qty(k', p'; \omega + \Omega) \, \bar{B}_{LL}\qty(p, k; \omega)+V^{\ast}_{k, k'} \, V^{\ast}_{p, p'} \,B_{LL}\qty(k', p'; \omega + \Omega) \, \bar{B}_{RR}\qty(p, k; \omega)\right]\,,\\
     & C_{o}\qty(\omega,\Omega) =\frac{-\pi}{2} \, \sum_{k, k', p, p'} \qty[1 - \tanh\qty(\frac{\omega + \Omega}{2 T}) \, \tanh\qty(\frac{\omega}{2 T})] \nonumber \\
     & \left[V_{k, k'} \, V^{\ast}_{p, p'} \, B_{RL}\qty(k', p'; \omega + \Omega) \, \bar{B}_{LR}\qty(p, k; \omega)+V^{\ast}_{k, k'} \, V_{p, p'} \,B_{LR}\qty(k', p'; \omega + \Omega) \, \bar{B}_{RL}\qty(p, k; \omega)\right]\,,
     \label{eq:ChiFourierSpec_odd}
\end{eqnarray}

\end{widetext}
which correspond to the electron-hole, even-$\omega$ and odd-$\omega$ contributions. In case of a weak link between two superconductors, the summations over momenta simplify and we can use the end-site Green's functions, and their corresponding spectral functions. We identify the terms containing $A_{\nu \nu'}$ as the electron-hole contribution. Finally, each contribution to the noise can be computed through Eq. (\ref{noise_def}), where the total noise will be the sum of the three contributions in Eqs. (\ref{eq:ChiFourierSpec_ph}-\ref{eq:ChiFourierSpec_odd}).


\end{document}